\documentclass{elsart}
\input epsf
\usepackage{epsfig}

\begin{document}
\begin{frontmatter}
\title{Detection of Minimum-Ionizing Particles and Nuclear Counter
Effect with Pure BGO and BSO Crystals with Photodiode Read-out}

\author[taiwan]{K.~Ueno\thanksref{email1}},
\author[taiwan,kek]{S.K.~Sahu\thanksref{email2}},
\author[taiwan]{K.C.~Peng},
\author[taiwan]{W.S.~Hou},
\author[nlt]{C.H.~Wang}

\address[taiwan]{National Taiwan University, Taipei, Taiwan, China}
\address[nlt]{National Lien Ho College of Tech. and Commerce, 
Miao Li, Taiwan, China}
\address[kek]{National Laboratory for High Energy Physics, KEK, 
Tsukuba, Ibaraki, Japan}

\thanks[email1]{ueno@kekvax.kek.jp} 
\thanks[email2]{Present address : University of Hawaii, Honolulu, 
HI 96822, U.S.A.; sahu@uhhepi.phys.hawaii.edu}

\begin{abstract}
Long BGO (Bi$_4$Ge$_3$O$_{12}$) and BSO (Bi$_4$Si$_3$O$_{12}$) crystals 
coupled with silicon photodiodes 
have been used to detect minimum-ionizing particles(MIP).
With a low noise amplifier customized for this purpose,
the crystals can detect MIPs with an excellent signal-to-noise
ratio. The {\em nuclear counter effect} is also clearly observed and 
measured. Effect of full and partial wrapping of a reflector 
around the crystal on light collection is also studied.
\end{abstract}


\end{frontmatter}

\sloppy
\baselineskip=0.8cm
\newpage
\section{Introduction}

Bismuth Germanate ($\mbox{Bi}_4\mbox{Ge}_3\mbox{O}_{12}$) crystals,
commonly known as BGO, have been extensively used for
electromagnetic(EM) calorimetry in high energy
physics experiments\cite{l3,topaz,cmd2}.
Advantages of BGO are its excellent $e/\gamma$ energy
resolution (0.3 -- 1 \%/$\sqrt(E(GeV)$),
high density(7.1 gm/cc), short radiation length(1.12 cm),
large refractive index(1.251), suitable scintillating properties(fast decay
time of about 300 $ns$ and peak scintillation at about 480 $nm$)
and non-hygroscopic nature. It is therefore
one of the best candidates for EM calorimetry in 
collider experiments,
especially where space imposes a serious constraint. 

Bismuth Silicate ($\mbox{Bi}_4\mbox{Si}_3\mbox{O}_{12}$) crystals,
known as BSO, on the other hand, although known to the 
particle physics community for some time\cite{koba}, are yet to
find a major deployment in a particle detector experiment. BSO has 
very similar properties as BGO~: high density(6.8 gm/cc), 
short radiation length(1.2 cm),
large refractive index(2.06), decay time of about 100 $ns$, peak
scintillation at about 450 $nm$ and non-hygroscopic nature.
Although it sells at about the same price as BGO at the moment, it has 
the advantage of being cheap if {\em commercially} produced, 
since the expensive
raw material germanium in BGO is replaced by silicon, which is 
much cheaper.  The light output of pure BSO crystal, however, is 
only about one-fourth of that of pure BGO, and hence energy resolution
of a calorimeter made up of BSO will be consequently worse than that
of a BGO calorimeter with similar geometry.

Both, pure BGO and pure BSO are known to be radiation hard at megarad 
level~\cite{koba,yano,sahu}, even upto 100 MRad.  This fact,
reinforced with the qualities cited in the above two paragraphs
makes these crystals potential materials for making 
high resolution EM calorimeters
at small angles(below 10$^\circ$) in B-factories.
Radiation level at such small angles
is rather high in B-factories due to intense flux of photons
and electrons generated by Bhabha events and spent-electron
background events\cite{tom}.
Such a calorimeter has been proposed, for example, for the BELLE
detector\cite{BELLE} at KEK B-factory to cover very small angles around
the beam pipe\cite{efc}. 

With such calorimeters one desires, driven by several physics
motivations, to detect not only 
EM showers, but also to tag minimum-ionizing
particles (MIP), such as high energy charged pions, muons, kaons
and protons.  Light output of these crystals at these 
small angles is typically read out by photodiodes(PD). Use of 
photomultiplier tubes(PMT) at such small angles is severely 
restricted due to issues like lack of space and high non-uniform 
magnetic field. PDs have much lower gain and much worse
signal-to-noise ratio than PMTs, and therefore detecting MIPs
becomes a challenge with a BGO(BSO)+PD system, since 
MIPs produce a lot less light than $e^\pm$ and $\gamma$.

In this paper we report our successful effort in 
making a low-noise amplifier system, with which we
can detect MIPs with BGO(or BSO)+PD system.
Although the light output of BSO is only $1/4$ of that of BGO,
we are still able to see the MIPs with this system.  
 
We describe the design and performance of the preamplifiers 
we developed for this purpose in the next section.
Setup for MIP detection by BGO and BSO crystals by 
using a high energy pion beam
and observation of nuclear counter effect is described in the third section. 
In the fourth section we make an analysis of the data and study the 
effect of reflector around 
the crystal on light collection.  Results are summarized 
in the last section.

\section{Amplifier}

A customized charge preamplifier was developed  
for amplification of signal from photodiodes.
Circuit diagram of the preamp is shown in Fig.~\ref{fig:circuit}. 
We adapted the design from a preamplifier
used in AMY experiment\cite{AMY} and experimented with different JFETs.
We settled on two brands of preamp : one using 2SK291, and the other using
2SK715 JFET as the main amplifying element. \\

\noindent
{\bf Calculation of gain~:} An input square pulse sequence of width 200
$\mu$s, and amplitude 60 mV was delivered into the TEST
key of the preamp (see Fig.~\ref{fig:circuit}). The output
of the preamp was digitized with a CAMAC ADC LeCroy 2249A. 
ADC counts are plotted in Fig.~\ref{fig:amp}(a). The peak at the 
left corresponds to the pedestal, whereas the peak on the right
gives the integrated charge. From the ratio of this peak 
(calibrated as 65 mV) to the charge input(10.5 fC), 
the gain of the preamp was calculated to be 6.2 V/pC. \\

\noindent
{\bf Test with Radioactive Source :} 
We tested two kinds of
photodiodes from Hamamatsu Photonics\cite{hama}: S5106 and S2662-03.
Active area and capacitance of S5106(S2662-03) are 5$\times$5 mm$^2$
(7.5$\times$20 mm$^2$) and 40 pF(100 pF), respectively.
The preamp was now coupled to the PD at the point shown in 
Fig.~\ref{fig:circuit}. 
An $^{241}$Am radioactive source was mounted on the PD, and the whole 
setup was placed in a light-tight box.
The source has a $\gamma$--ray peak at 60 keV, which may sometimes be 
absorbed completely without any energy leakage by the 300 $\mu$m thick 
depletion layer of the photodiode.

Signals generated by these $\gamma$--rays in the PD were 
amplified in the preamp, self-triggered, and integrated by a
CAMAC ADC LeCroy 2249A with a gate width of 200 ns.

The 60 keV peak is easily seen with the system.  The corresponding 
pulse-height spectra for different combinations of photodiodes
and JFET are shown in Fig.~\ref{fig:amp}(b), (c) and (d).
These figures are for different PDs and different JFETs, as
indicated in the respective plots.

Fits to the 60 keV peak and pedestal for all these 
three combinations of PD and JFET are given in Table~1. \\

\noindent
{\bf Calculation of ENC :} While working with photodiodes,
one often wants to know the noise or resolution of the system
in terms of electron-hole pairs produced in the PD. 
{\em Equivalent Niose Charge} or ENC represents such a 
measure. In Fig.~\ref{fig:amp}(b), for example, width of 
the peak is 3.29\% of the mean(Table~\ref{tab:tab1}), and hence the noise is 
3.29\% of the signal produced. Since the energy required 
to produce an electron-hole pair in silicon is about 
3.6 eV, a 60 keV photon produces 16,667 electrons in the PD.
So the noise translates to 16,667 $\times$ 3.29\% = 548 electrons,
which is the ENC for this system. ENC for the other two systems 
are 970 and 906, as posted in Figs.~\ref{fig:amp}(c) and (d), respectively.

It is clear that the configuration of \underline{PD S5106} and preamp with 
\underline{JFET 2SK715} renders the least ENC, and hence corresponds to 
least noise. We therefore \underline{chose this system} to measure 
the scintillation of BGO and BSO crystals. \\

\noindent
{\bf Estimation of S/N for MIP :} A MIP deposits an energy of 
about 100 MeV in the length of our BGO crystal.
About 300 eV is needed for one scintillation in pure BGO.
Assuming about 20\% light collection efficiency and 100\%
quantum efficiency of PD, we end up with about 
66,000 electron-hole pairs created in the PD for a MIP.
Since the ENC is 548 electrons, we can expect a signal-to-noise
ratio (S/N) of about 120~:~1 with our system. For BSO, since the 
light output is about one-fourth of BGO, the S/N would be 
about 30~:~1.

\section{BGO and BSO Crystals on MIPs}

We experimented on three samples with the same cross-sectional area of 
1$\times$1 cm$^2$: 
(A) 10 cm long BGO crystal from the Institute of Inorganic 
Chemistry, Novosibirsk, Russia\cite{novo},
(B) 12 cm long BGO crystal from the Institute of Single Crystal,
Ukraine\cite{ukra}, and 
(C) 12 cm long BSO crystal from Futec Furnace Co, Japan\cite{fute}.
Photodiodes(S5106) were glued to
one end of crystals with an optical glue called Eccobond\cite{eccobond}.
The crystals were then wrapped hermetically,  
first with 150 $\mu$m teflon tapes 
for better light collection and then with black tapes, 
for protection against light leak from outside.  

The samples were exposed to a 3.5 GeV $\pi^-$ beam at the $\pi 2$ 
beam line at the KEK-PS.  A schematic diagram of the set-up 
is given in Fig.~\ref{fig:layout}. 
The $e/\pi$ separation in the beam was achieved with a 
CO$_2$ \v{C}erenkov counter.
The trigger was provided by the coincidence of three scintillation 
counters along the beam.
The pions would then enter the volume of the crystal (and sometimes
pass through the PD, too), and deposit some energy to produce the
scintillation light, which would then be collected by the 
photodiode.  Signal from the PD was amplified by the 
preamp, and was 
digitized by a CAMAC ADC LeCroy 2249W module with 
a  4 $\mu$sec gate initiated by the trigger. The data was logged 
by a Unix workstation-based DAQ system.
The crystal glued with PD and preamp were placed in a 
light-tight box made up of thick aluminum, which was electrically
grounded, and therefore served as an excellent Faraday cage.
The pedestal was
logged concurrently by triggering the DAQ with a clock of the same 
gate width, asynchronous with the beam gate.

\section{Results and Analysis}

\noindent
{\bf MIP Detection~: } ADC spectra for 
the three samples A, B and C for the set-up
described above are shown in Figs.~\ref{fig:mipsignal}(a),
(b) and (c), respectively.
The first peak in each spectrum corresponds to the pedestal,
the second one to the MIP, and the third one to the sum of the MIP and 
the Nuclear Counter Effect(NCE)\footnote{The Nuclear Counter Effect(NCE)
is the extra amount of charge produced in the photodiode 
by a charged particle directly hitting it, on the top of the 
charge produced by the scintillation light. A MIP, for example,  produces 
about 25000 electron-hole pairs in a photodiode of thickness 
300 $\mu$m.  Nuclear counter effect worsens the resolution 
of an EM calorimeter, where some of the secondary $e^\pm$
might hit the PD. This effect is avoided by using enough radiation
lengths of crystal along the direction of the shower and/or using
{\em avalanche photodiodes}.
For MIP detection, however, it does not pose a problem as long as 
the resulting signal due to NCE is comparable or 
less than the MIP scintillation
signal, which is the case in this experiment.}.
The third peak thus 
corresponds to the event where the pion deposited energy along the
length of the crystal, and then hit the photodiode.  The difference 
in the second and the third peak corresponds to the amount of energy
deposited in the photodiode itself when a minimum ionizing pion traverses it.
This conjecture was confirmed by a simple calculation of energy loss and
a GEANT\cite{Geant} simulation indicated by the dashed line in 
Fig.~\ref{fig:mipsignal}(a). The simulation is not normalized to 
the real data, to retain the clarity of the comparison.

In Fig.~\ref{fig:mipsignal}(d) we show the ADC logged when 
the crystal is removed from the setup, {\em i.e.}, when the 
beam directly hits the PD. Difference in ADC counts 
between the two peaks corresponds to the signal generated 
by the NCE.  It may be noted that this 
difference is same as the difference between second and third 
peaks in Figs.~\ref{fig:mipsignal}(a),(b) and (c), which 
confirms that the third peak in these figures corresponds to 
the nuclear counter effect indeed.

It is also apparent from Fig.~\ref{fig:mipsignal} that 
sample A has about 40\% more light
output than B after correcting for the length, which is not
surprising since they are from different manufacturers, and 
BGO light output is known to be quite sensitive to production
method and trace impurities.  The BSO sample C
has about 25\% light output compared to BGO sample A, as
already observed in Ref.~\cite{koba}, and one is still
able to observe the MIP peak. \\

\noindent
{\bf Effect of reflector~: }We also studied the advantage 
of the teflon reflector around the crystal.
BGO has a high refractive index of 2.15, and therefore is supposed to 
retain most of the scintillation light by total internal
reflection. We did the following experiment in order to
study the effect of putting on a teflon reflector 
around the crystals. First, we stripped the reflector off the 
sample B except for the very end opposite to the 
photodiode(we will call this setup as ``BGO with a reflector-cap''). 
Then the sample was subjected to the 3.5 GeV $\pi^-$ beam.
The pulse height spectrum is given in
Fig.~\ref{mipsignal_reflector}(a)
as the solid line.  Then we removed the 
reflector completely, leaving the crystal bare. 
It was then subjected to the beam,
and the observed pulse-height spectrum is shown in 
Fig.~\ref{mipsignal_reflector}(a) as the dotted line.

We can easily distinguish the MIP and nuclear counter peaks in these
two superimposed plots. It can be readily seen by comparing the
position of the two MIP peaks after pedestal subtraction, that
a reflector cap increases the light collection in BGO by about 30\%
compared to a bare crystal.

We then took the sample A, completely wrapped with the reflector,
and subjected it to the beam. The ADC spectrum is plotted 
in Fig.~\ref{mipsignal_reflector}(b) as the solid line. Then we 
stripped the reflector completely off the sample, and repeated 
the experiment. The corresponding ADC spectrum is plotted as the 
dotted line in Fig.~\ref{mipsignal_reflector}(b).

Again, the peaks due to the MIP and NCE 
are clearly visible. By comparing the two MIP peaks after 
pedestal subtraction, it can be seen that the light collection 
with full reflector wrap improves by about 85\% compared to the
bare crystal.

It is interesting to note that (Fig.~\ref{mipsignal_reflector})
the height of the NCE
peak with respect to the MIP peak is smaller for sample B
compared to that for sample A. The reason may be ascribed 
to the larger length of sample B, where more scintillation light is 
collected by the PD, thereby pulling down the ratio.
 

\section{Summary}

\begin{itemize}

\item BGO and BSO crystals coupled with photodiodes are proven to be
capable of detecting minimum-ionizing particles
with a large S/N ratio.  
The preamplifier used is perfectly adequate for the purpose.
The signal of MIPs is well separated from electronic noise and 
NCE signal.

\item The detection of MIPs with BSO coupled with a PD is reported 
for the first time.

\item A clear effect of NCE in a calorimetric 
environment is reported for the first time.

\item Effect of reflector wrap around the crystal in regards to the 
light collection is studied.

\end{itemize}

\section*{Acknowledgements}
The MIP detection experiment was done under the auspices of National Lab for 
High Energy Physics (KEK) as the experiment T-388 of the KEK-PS.
We would like to thank  Dr.~M.~Kobayashi for providing valuable
information on BSO, and Dr.~Y.~Sugimoto for valuable suggestions on
preamps. 
The Aerogel and CsI subgroup members of the BELLE collaboration
have been extremely helpful in this project.
This experiment was supported in part by the grant NSC~85-2112-M-002-034
of the Republic of China. 


\newpage
\section*{Table Captions}
{\bf Tab. 1 : }Peak channels and widths for Fig.~2, see text for details.

\section*{Figure Captions}
\noindent
{\bf Fig. 1 : }Circuit diagram of the charge preamplifier.
Shaping time is about 1 $\mu$s. 

\noindent
{\bf Fig. 2 : }Study of the charge preamplifier~:
(a) Response of preamp to a test pulse. The charge-amp gain is 
measured to be 6.2 V/pC;
(b) Response of PD(S5106) + preamp with JFET(715) to 60 keV 
$\gamma$-rays from $^{241}$Am. ENC is 548 electrons;
(c) Same configuration as (b), but with a larger PD(S2662-03).
ENC is 970 electrons;
(d) Same configuration as (c) but JFET switched to 2SK291. 
ENC is 906 electrons.

\noindent
{\bf Fig. 3 : }Setup for the MIP detection experiment.

\noindent
{\bf Fig. 4 : }Observation of MIP signal with 3.5 GeV $\pi^-$ beam~:
(a) Observed ADC counts for 10 cm long BGO sample A. 
Solid line is the real data, and 
dashed line is simulated data(not normalized with real data).
(b) Observed ADC counts for 12 cm long BGO sample B.
(c) Observed ADC counts for 12 cm long BSO sample C.
(d) Observed ADC counts for MIP hitting PD directly in the 
absence of any crystal.

\noindent
{\bf Fig. 5 : }Study of the effect of Teflon reflector on light collection~:
In the top, the three configurations are sketched, and referred to 
in the plots.
(a) Observed ADC counts for sample B, with complete reflector and 
with partial reflector.
(b) Observed ADC counts for sample A, with complete reflector and 
with no reflector.

\newpage
\begin{center}
\begin{table}[h]
\begin{tabular}{|c||r@{$\ \pm\ $}l|r@{$\ \pm\ $}l||r@{$\ \pm\ $}l|}
\hline
\hline
Condition & \multicolumn{2}{c|}{Signal Peak} & 
\multicolumn{2}{c||}{Gaussian $\sigma$} & \multicolumn{2}{c|}{Pedestal} \\
\hline
Test Pulse &  1038. & .04 & 7.6 & .04 & 24.4 & .03 \\ \hline
Small PD(S5106) + JFET(715) & 507.5 & .1 & 15.9 & .2 & 23.5 & .06 \\ \hline
Large PD(S2662-03) + JFET(715) & 489.3 & .3 & 27.2 & .3 & 22.0 & .2 \\ \hline
Large PD(S2662-03) + JFET(291) & 459.0 & .2 & 23.8 & .3 & 21.2 & .2 \\ \hline
\hline
\end{tabular}
\caption{ Peak channels and widths for Fig.~2, see text for details.}
\label{tab:tab1}
\end{table}
\end{center}

\begin{figure}[h]
\epsfig{file=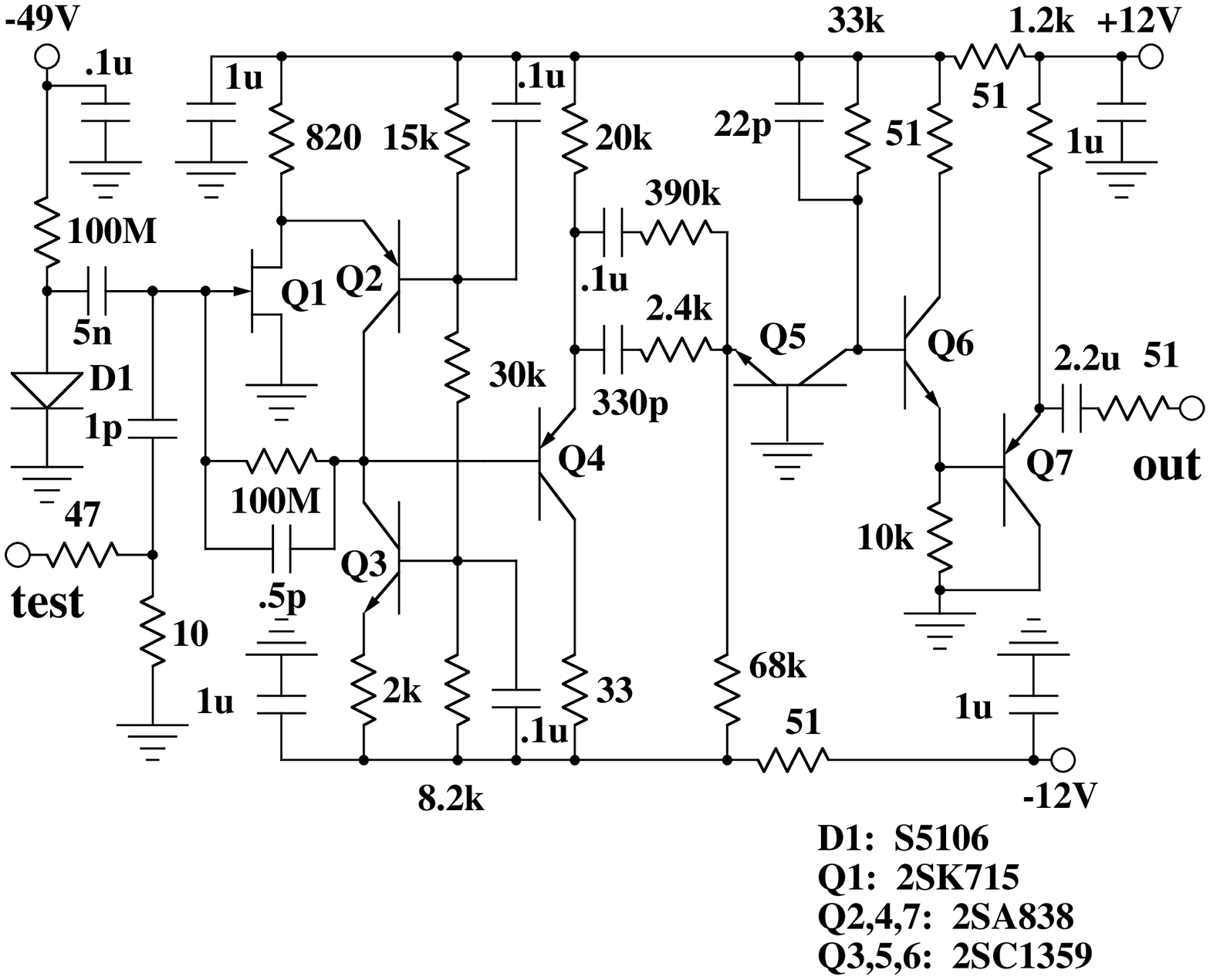,width=6.5in,height=5.in}
\caption{
Circuit diagram of the charge preamplifier.
Shaping time is about 1 $\mu$s.
}
\label{fig:circuit}
\end{figure}

\newpage
\section*{Figure \ref{fig:circuit}, K. Ueno et al., NIM-A}

\begin{figure}[h]
\epsfig{file=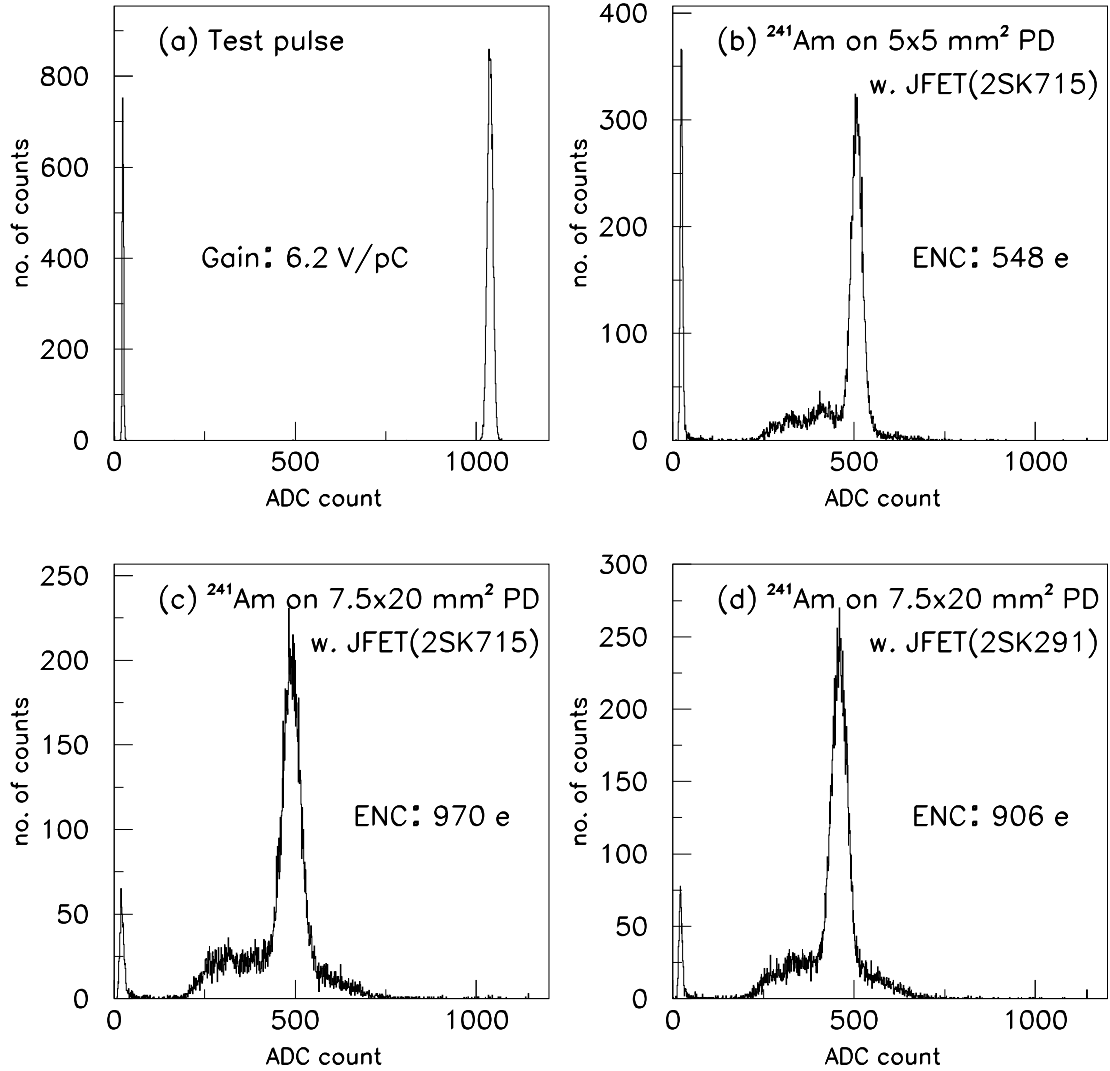,width=6.5in,height=6.5in}
\caption{
Study of the charge preamplifier~:
(a) Response of preamp to a test pulse. The charge-amp gain is 
measured to be 6.2 V/pC;
(b) Response of PD(S5106) + preamp with JFET(715) to 60 keV 
$\gamma$-rays from $^{241}$Am. ENC is 548 electrons;
(c) Same configuration as (b), but with a larger PD(S2662-03).
ENC is 970 electrons;
(d) Same configuration as (c) but JFET switched to 2SK291. 
ENC is 906 electrons.
}
\label{fig:amp}
\end{figure}

\newpage
\section*{Figure \ref{fig:amp}, K. Ueno et al., NIM-A}

\begin{figure}[h]
\epsfig{file=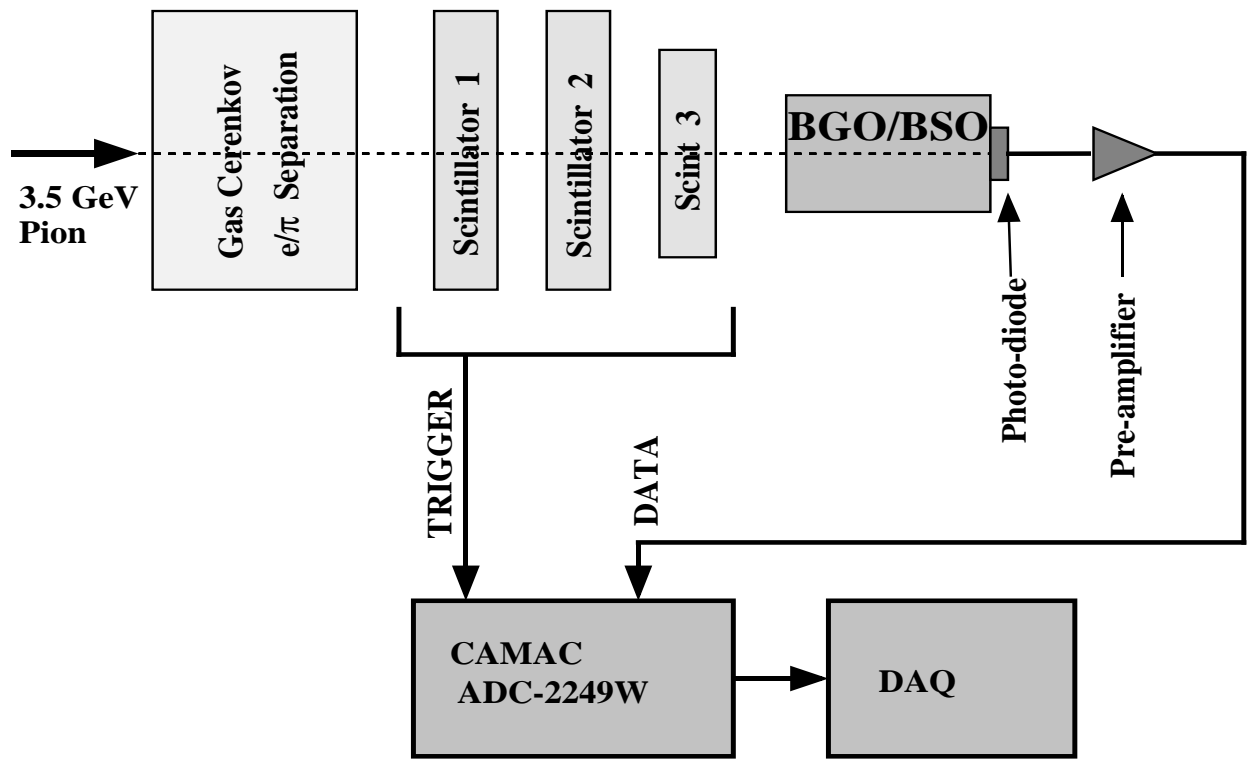,width=6.5in,height=4.0in}
\caption{
Setup for the MIP detection experiment.
}
\label{fig:layout}
\end{figure}

\newpage
\section*{Figure \ref{fig:layout}, K. Ueno et al., NIM-A}

\begin{figure}[h]
\epsfig{file=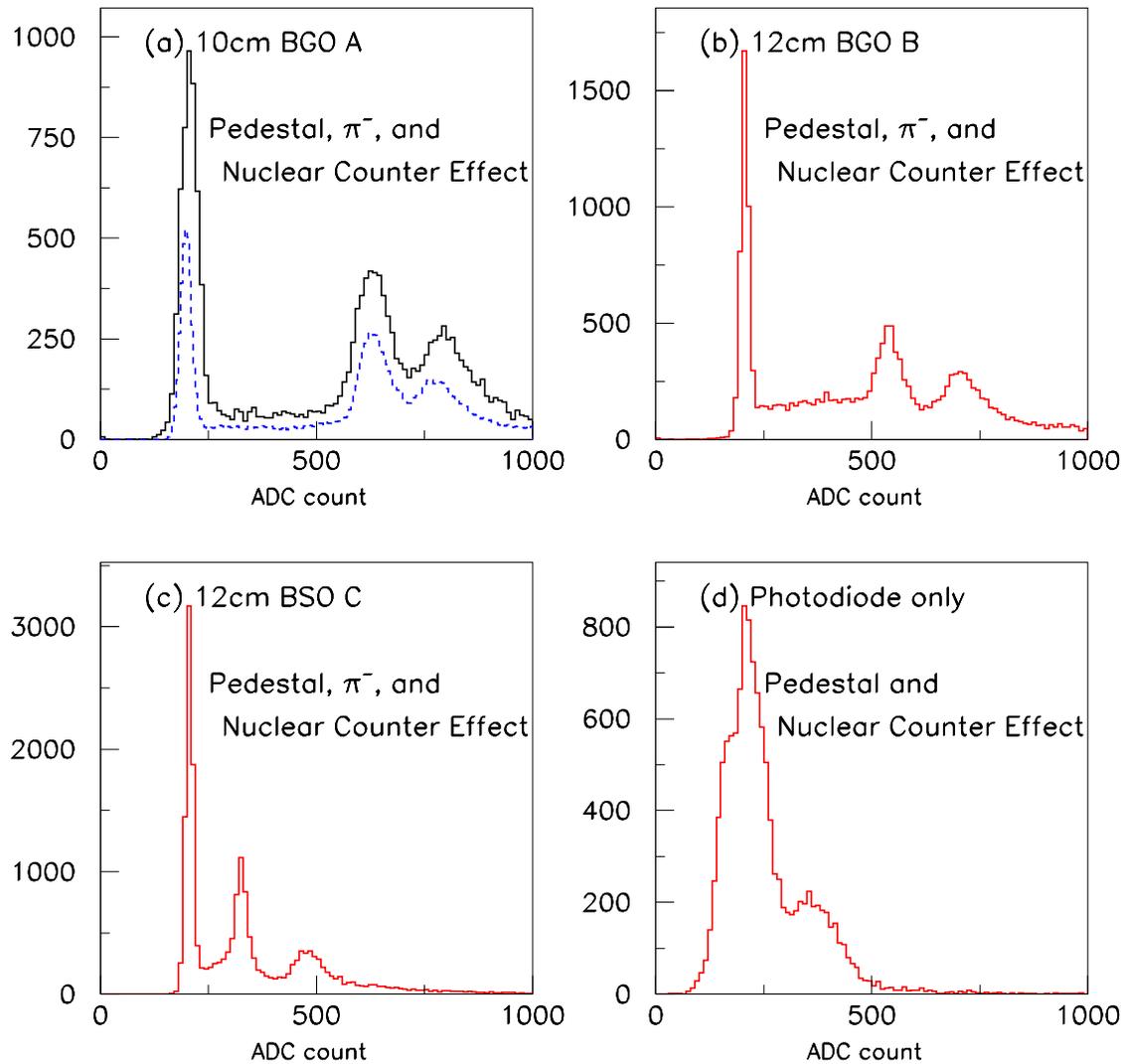,width=6.5in,height=6.5in}
\caption{
Observation of MIP signal with 3.5 GeV $\pi^-$ beam~:
(a) Observed ADC counts for 10 cm long BGO sample A. 
Solid line is the real data, and 
dashed line is simulated data(not normalized with real data).
(b) Observed ADC counts for 12 cm long BGO sample B.
(c) Observed ADC counts for 12 cm long BSO sample C.
(d) Observed ADC counts for MIP hitting PD directly in the 
absence of any crystal.
}
\label{fig:mipsignal}
\end{figure}

\newpage
\section*{Figure \ref{fig:mipsignal}, K. Ueno et al., NIM-A}

\begin{figure}[h]
\epsfig{file=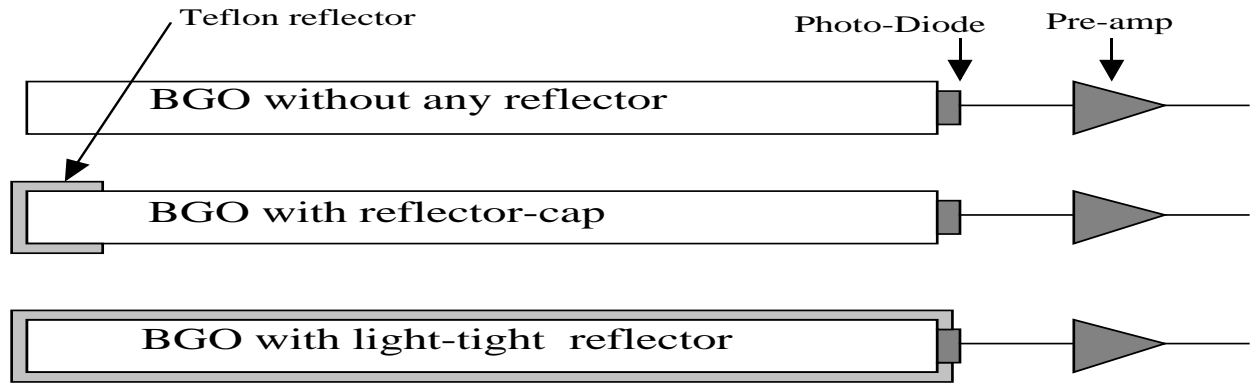,width=6.5in,height=2in}
\epsfig{file=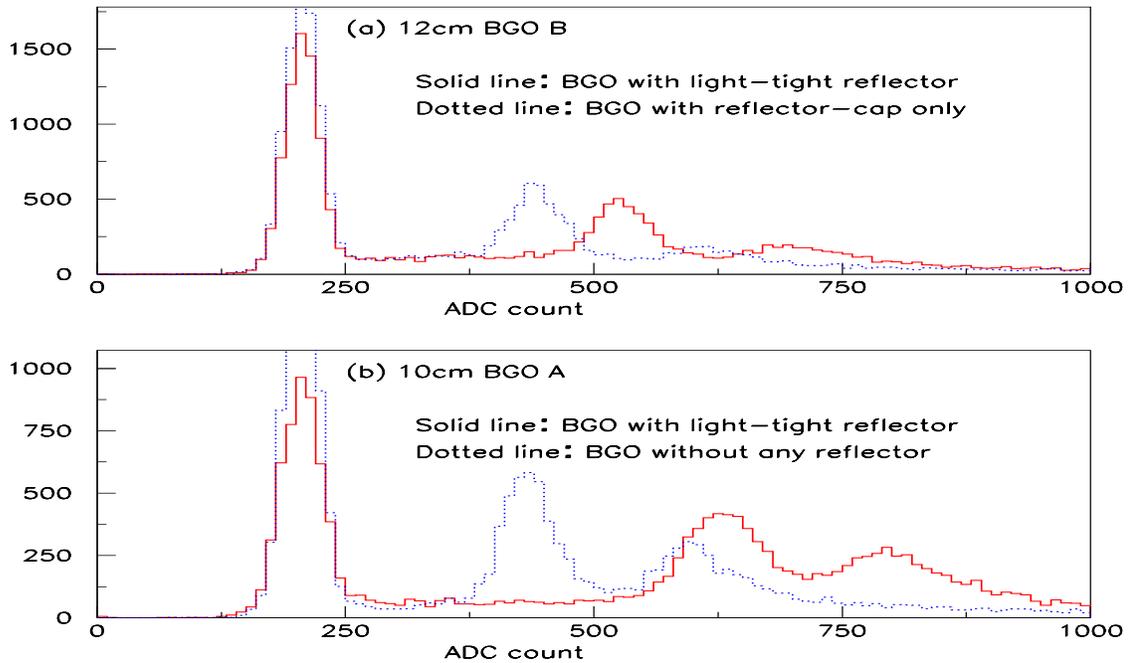,width=6.5in,height=4in}
\caption{
Study of the effect of Teflon reflector on light collection~:
In the top, the three configurations are sketched, and referred to 
in the plots.
(a) Observed ADC counts for sample B, with complete reflector and 
with partial reflector.
(b) Observed ADC counts for sample A, with complete reflector and 
with no reflector.
}
\label{mipsignal_reflector}
\end{figure}

\newpage
\section*{Figure \ref{mipsignal_reflector}, K. Ueno et al., NIM-A}

\end{document}